\documentclass[11pt]{article}
\usepackage{epsfig,url}
\usepackage{setspace}
\usepackage{geometry}
\begin{document}

\doublespace

\title{Chinese Internet AS-Level Topology}

\author{{ Shi~Zhou}\\Department of Computer Science, University~College~London
\\Ross Building, Adastral Park, Ipswich, IP5 3RE, United Kingdom\\email: s.zhou@adastral.ucl.ac.uk\\~\\
{ Guo-Qiang~Zhang and Guo-Qing~Zhang} \\Institute of Computing
Technology, Chinese
Academy of Sciences\\Ke-xue-yuan South Road, Beijing, 100080, China\\
email: guoqiang@ict.ac.cn, gqzhang@ict.ac.cn}

\maketitle

\begin{abstract}

We present the first complete measurement of the Chinese Internet
topology at the \emph{autonomous systems} (AS) level based on
traceroute data probed from servers of major ISPs in mainland China.
We show that both the Chinese Internet AS graph and the global
Internet AS graph can be accurately reproduced by the
Positive-Feedback Preference (PFP) model with the same parameters.
This result suggests that the Chinese Internet preserves well the
topological characteristics of the global Internet. This is the first
demonstration of the Internet's topological fractality, or
self-similarity, performed at the level of topology evolution
modeling.

\end{abstract}



\section{Introduction}

During the last three decades, the Internet has experienced
fascinating evolution, both exponential growth in traffic and
endless expansion in topology. It is crucial to obtain a good
description of the topology, because effective engineering of the
Internet is predicated on a detailed understanding of issues such
as the large-scale structure of its underlying physical topology,
the manner in which it evolves over time, and the way in which its
constituent components contribute to its overall
function~\cite{floyd03}.

The Internet contains millions of routers, which are grouped into
thousands of subnetworks, called \emph{autonomous systems} (AS).
Network structure inside an AS only affect local properties, whereas
the delivery of data traffic through the global Internet depends on
the complex interactions between ASes that exchange routing
information using the Border Gateway Protocol (BGP)~\cite{quoitin03}.
This paper focuses on the Internet AS-level topology, on which a node
is an AS and a link represents a BGP peering relation between two
ASes. Measurements of the Internet AS-level topology~\cite{nlanr,
oregon, caida} became available in late 1990's. These measurements
have greatly improved our understanding of the evolution and
structure of the Internet~\cite{pastor04}.

During the last decade, the Internet has developed rapidly in China.
However measurement of the Chinese Internet topology has only been
carried out from outside of the country. In this paper, we present
the first and so far the most complete measurement of the Chinese
Internet AS graph based on traceroute data probed from servers of
major ISPs inside mainland China. The obtained Chinese Internet AS
graph is a small regional subgraph of the massive global Internet AS
graph because it contains only ASes located in mainland China. We
show that both the global graph and the small local graph can be
accurately reproduced by the Positive-Feedback Preference (PFP)
model~\cite{zhou04d, zhou06b} with the same parameters. This result
suggests that the Chinese Internet preserves well the topological
characteristics of the global Internet. Our work demonstrates the
existence of a statistical fractality in the description of the
Internet AS-level topology.

\section{Measurements of Internet Topology}

There are two primary methods for inferring Internet AS-level
topology: one is the passive measurements inferring AS connectivity
information from BGP routing tables, e.g.~\cite{nlanr, oregon}; and
the other is the active measurement which extracts AS graph from
traceroute probing data, e.g.~\cite{caida}. AS graphs produced by the
two methods exhibit a number of different topology
properties~\cite{mahadevan05b}. The BGP measurements reflect the
control plane of the AS topology, whereas the traceroute measurements
reflect the data plane of the AS topology. BGP tables have the
advantage that they are relatively easy to parse, process and
comprehend, however BGP measurements suffer certain limitations in
that they do not reflect how traffic \emph{actually} travels toward a
destination network and they rarely include Internet exchange points
(IXP) ASes. Traceroute measurements also have inherent limitations
associated with converting IP addresses (on the router level) to AS
numbers, however traceroute measurements reflects the \emph{actual}
paths that IP packets follow. In this paper we focus on the active
measurements based on traceroute data.

\subsection{The Global Internet AS Graph -- \texttt{ITDK0304}}

In 1998, CAIDA~\cite{caida} launched the Macroscopic Topology Project
to collect and analyze Internet-wide topology at a representatively
large scale. CAIDA's topology measurement tool, \emph{skitter},
implements the Internet Control Message Protocol (ICMP). It collects
the forward path from the monitor to a given destination and captures
the addresses of intermediate routers in the path. The {skitter} runs
on more than 20 monitors around the globe and actively collects
forward IP path to over half a million destinations across the
current IPv4 address space. CAIDA extracts interconnect information
of ASes from the massive traceroute data collected by {skitter} and
produces AS adjacency graphs. CAIDA argues~\cite{hyun03} that while
inherent limitations of the traceroute-based probing methodology do
not allow for 100\%-accurate extraction of the real Internet topology
from skitter data, CAIDA seeks data sources that are collectively
most likely to capture a precise and coherent snapshot of macroscopic
Internet structure. We extract the global Internet AS graph in this
paper from the Internet topology data kit {\texttt{ITDK0304}}
collected by CAIDA in April 2003~\cite{itdk0304}. The ITDK0304
Internet AS graph contains 9024 nodes and 28959 links.

\subsection{The Chinese Internet AS Graph -- \texttt{CN05}}

The Internet in China  has developed rapidly in the last decade.
Today the country has more than 100 million Internet
users~\cite{CNNIC18}. However measurement of the Chinese Internet
topology has only been carried out from outside of the country. For
example, currently there is no skitter monitor located in China.

Recently we measured the Chinese AS graph  using the active
measurement method. We collected traceroute data from six servers
belonging to the following major ISPs in mainland China: CSTNET
(China Science and Technology Net), CERNET (China Education and
Research Net), CAPINFO Company Limited, ChinaTelecom, ChinaNetCom and
CGWNET (China Great Wall Net). The measurement process took place
during the first week in May 2005.

From each server, a traceroute tool probes 7436 destination IP
addresses in mainland China, including (i) 3010 Web server addresses,
which are obtained from the Chinese Web yellow pages
database~\cite{yellowpage}, and (ii) 4426 random IP addresses, which
are obtained by using the stub network sampling method~\cite{jiang06}
which samples one IP address from every IP prefix assigned to
mainland China by APNIC~\cite{apnic}. The traceroute tool probes each
IP address for three times. Although all the Web server addresses are
reachable by HTTP, they are not 100\% reachable by the traceroute
tool because many Web servers and ASes do not allow ICMP probing from
external servers. Of the 3010 Web server addresses, the traceroute
tool reached 1859 addresses and ASes to which 2753 addresses belong.
Of the 4426 random IP addresses, the traceroute tool reached 336
addresses and ASes of 3485 addresses. The success rate of probing
random IP addresses is even lower because many randomly sampled IP
addresses do not exist. We observed that the size of obtained Chinese
AS graph will not increase significantly when probing more
destination addresses.

We accumulated all the traceroute data and used the same techniques
of CAIDA~\cite{hyun03} to convert IP addresses to corresponding AS
numbers announced by APNIC. The resulted Chinese AS graph is called
\texttt{CN05}, which contains 84 nodes and 211 links. The dataset is
available at \url{http://www.adastral.ucl.ac.uk/~szhou/resource.htm}
in the format of a list of peering AS numbers. Our measurement is
more complete than the Chinese AS graph extracted from CAIDA's
ITDK0304 which contains only 71 Chinese ASes and 160 peering links
among them.

\section{Comparison Between The Two AS Graphs}

The Chinese Internet AS graph CN05 is a small subgraph of the global
Internet AS graph ITDK0304. In this paper we compare the two AS
graphs in an indirect way, i.e.,~we contrast the two AS graphs
against networks generated by the Positive-Feedback Preference (PFP)
model~\cite{zhou04d,zhou06b}, which has been regarded by CAIDA as the
most precise and complete Internet topology generator to
date~\cite{rrfs05, nsf04-540, krioukov06}.

\subsection{The Positive-Feedback Preference Model} The PFP model is an extensive
modification of the Barab\'asi-Albert model~\cite{barabasi99a}. The
PFP model grows Internet-like networks by using the following two
generative mechanisms, namely the interactive growth and the
positive-feedback preference, which are coincident with a number of
observations on the Internet history
data~\cite{chen02,park03,vazquez02}.

\begin{figure}[tbh]
\centerline{\psfig{figure=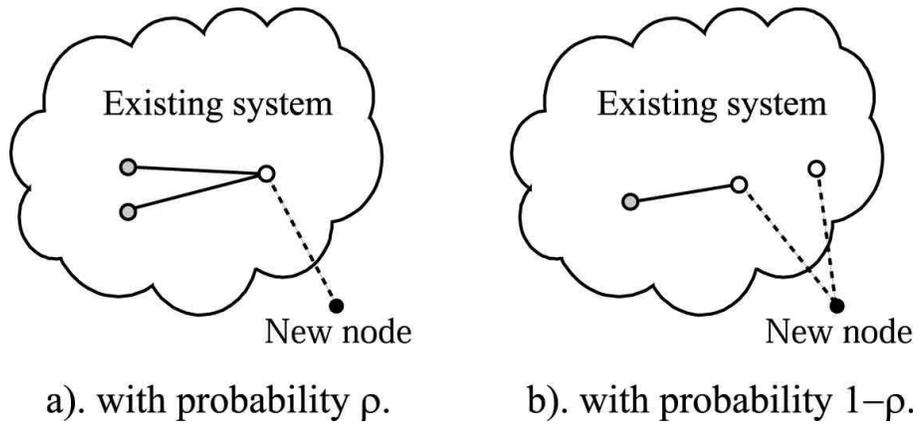,width=12cm}}
\caption{\label{fig:Growth} Interactive growth of the PFP model. }
\end{figure}

\paragraph{Interactive Growth.}\label{section:IGmodel}Starting from a small random graph, at each time step,
1) with probability $p\in[0, 1]$ (see Fig.\,\ref{fig:Growth}a), a
new node is attached to an old node in the existing system, and
then two new links are added connecting the old node to two other
old nodes; and 2) with probability $1-p$ (see
Fig.\,\ref{fig:Growth}b), a new node is attached to two old nodes,
and then one new link is added connecting one of the two old nodes
to another old node.

\paragraph{Positive-Feedback Preference.} Degree $k$ is defined as the number of links a node has.
The preference probability that a new link is attached to node $i$
is given as
\begin{equation} \Pi(i) =
{k_i^{1+\delta\log_{10}{k_i}}\over\sum_j
k_j^{1+\delta\log_{10}{k_j}}},~~\delta\in[0, 1], \label{eq:PFP}
\end{equation}
in which a node's ability of acquiring new links increases as a
feed-back loop of the node degree. Such a preference scheme results
in ``the rich not only get richer, but get disproportionately
richer''.

Numerical simulation shows that the PFP model produces the best
result when the parameters $p=0.4$ and
$\delta=0.048$~\cite{zhou04d,zhou06b}. As stated in the status report
of IRTF Routing Research Group's (RRG) Future Domain Routing (FDR)
Scalability Research Subgroup (RR-FS) in 2005~\cite{krioukov06}, the
PFP  model ``is the best non-equilibrium network growth model, with
respect to its proximity to observed Internet topology.''

\subsection{Simulation Result}
We use the PFP model to grow networks to the same number of nodes as
CN05 and ITDK0304 (see Table~\ref{table:properties}) using the same
model parameters as suggested in previous numerical
simulations~\cite{zhou04d,zhou06b}, i.e.,~$p=0.4$ and $\delta=0.048$.
For each AS graph, ten networks are generated using random seeds and
their properties are averaged.

\begin{table}
\caption{AS Graphs vs PFP Networks.} \label{table:properties}
\centering
\renewcommand{\tabcolsep}{1.2pc} 
\renewcommand{\arraystretch}{1.5} 
\begin{tabular}{c|cc|cc}
\hline\hline
~&             CN05 & PFP$_{1}$ & ITDK0304 & PFP$_{2}$ \\
\hline
Number of nodes $N$ & 84& 84& 9204&  9204 \\
Number of links $L$ & 211& 217& 28959& 27612 \\
Maximum degree $k_{max}$    & 38& 39 & 2070&1950 \\
Degree exponent $\gamma$  &-2.21&-2.21&    -2.254 & -2.255\\
Assortative coefficient $\alpha$    &-0.328&-0.298&-0.236& -0.234 \\
Top clique size $n_{clique}$ & 3 & 3 & 16& 16\\
Rich-club exponent $\theta$ &-1.42&-1.42 &-1.48 & -1.49\\
Average triangle coefficient& 5.6 & 6.8 & 21.4 & 19.1\\
Characteristic path length $\ell^*$ &2.54&2.54&3.12&3.07\\
Maximum coreness $c_{max}$ & 5 & 6 & 28 & 27\\
\hline\hline\end{tabular}
\end{table}

The Internet topology is extremely complex. Researchers have proposed
a number of metrics to quantify and explain the structure of
Internet~\cite{pastor04}. In this paper we compare the AS graphs and
the PFP networks by examining the following topological properties:
\begin{enumerate}
\item Degree distribution \item Degree correlation \item Rich-club
connectivity \item Triangle coefficient \item Shortest path length
\item $k$-core structure
\end{enumerate}
Note that CN05 is a very small graph, and therefore all its
statistics are of very low statistical significance.

\subsubsection{Degree Distribution}

Degree is a local property but the probability distribution of
degree provides a view on the global structure of a network.
Faloutsos\,{\sl et\,al}\,\cite{faloutsos99} showed that the
Internet AS-level topology  exhibits a power-law degree
distribution,
$$P(k) \sim k^\gamma, ~~\gamma\simeq-2.2.$$ This
means a few nodes have very large numbers of links, however the vast
majority of nodes have only a few links. This discovery is
significant because it invalidated previous Internet models that were
based on the classical random graphs having Poisson degree
distributions.

\begin{figure}[tbh]
\centerline{\psfig{figure=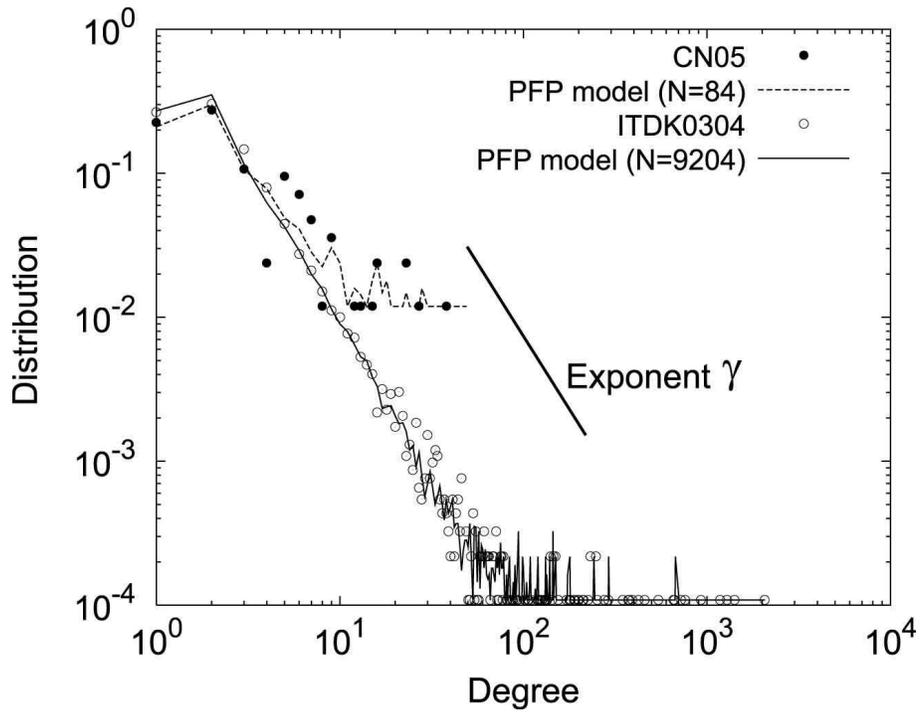,width=12cm}}
\caption{\label{fig:evaluation:PK}Degree distribution.}
\end{figure}

Fig.~\ref{fig:evaluation:PK} shows that ITDK0304 and CN05 are
characterized by a power-law degree distribution (Degree distribution
of CN05 is not statistically significant due to its small number of
nodes.) with the power-law exponent of CN05, $\gamma=-2.21$, is
slightly larger than that of the ITDK0304, $\gamma=-2.25$ (see
Table~\ref{table:properties}). It is clear that the PFP networks
closely match the degree properties of the two AS graphs.

As~\cite{tangmunarunkit02} pointed out, the degree distribution alone
does not uniquely characterize a network topology. To describe the
full picture of network structure, we need to examine other important
topological properties as well.

\subsubsection{Degree Correlation}

Recently a number of studies have shown that the degree correlation,
or the joint degree distribution $P(k, k')$, defined as the
probability that a link connects $k$- and $k'$-degree nodes, plays a
significant role in defining a network's
structure~\cite{mahadevan05b, newman03, vazquez03, maslov04}.
Networks exhibit different mixing patterns according to their degree
correlation. Social networks are assortative mixing where nodes tend
to attach to alike nodes, i.e.~high-degree nodes to high-degree nodes
and low-degree nodes to low-degree nodes. On contrast technological
and biological networks, including the Internet, are dissasortative
mixing where high-degree nodes tend to connect with low-degree nodes,
and visa versa.

A network's mixing pattern can be indicated by the correlation
between a node's degree $k$ and its nearest-neighbors average degree
$k_{nn}$~\cite{vazquez02}. A mixing pattern can also be identified by
measuring the assortative coefficient~\cite{newman03}, $-1<\alpha<1$,
which is defined as
$$
\alpha = {L^{-1}\sum_{m} j_m k_m -
[L^{-1}\sum_m{1\over2}(j_m+k_m)]^2 \over
L^{-1}\sum_m{1\over2}(j_m^2+k_m^2)-[L^{-1}\sum_m{1\over2}(j_m+k_m)]^2},
$$
where $L$ is the number of links a network has, and $j_m$, $k_m$ are
the degrees of the nodes at the ends of the $m$th link, with
$m=1,...,L$. If $0<\alpha<1$, a network is assortative mixing; and if
$-1<\alpha<0$, a network is dissasortative mixing.

\begin{figure}[tbh]
\centerline{\psfig{figure=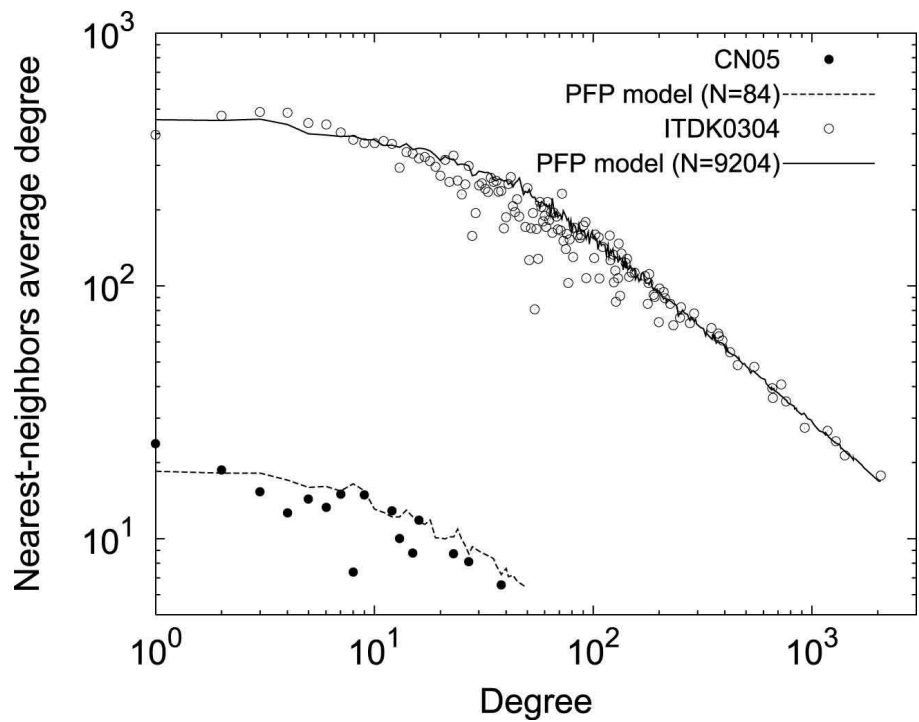,width=12cm}}
\caption{\label{fig:evaluation:Knn}Nearest-neighbors average degree
of $k$-degree nodes.}
\end{figure}

Fig.~\ref{fig:evaluation:Knn} shows that ITDK0304 and CN05 clearly
exhibit a negative correlation between degree and the
nearest-neighbors average degree. Also as shown in
Table~\ref{table:properties}, both the AS graphs are characterized by
a negative assortative coefficient. The PFP networks precisely
reproduce the disassortative mixing properties of the Internet AS
graphs.

\subsubsection{Rich-Club Connectivity}

A hierarchy property of the Internet's structure is the rich-club
phenomenon\,\cite{zhou04a}, which describes the fact that well
connected nodes, \emph{rich} nodes, are tightly interconnected with
other rich nodes, forming a core group or rich-club. The rich-club
phenomenon does not imply that the majority links of the rich nodes
are directed to other club members. Indeed, rich nodes have very
large numbers of links and only a few of them are enough to provide
the interconnectivity to other club members, whose number is anyway
small.

The rich-club membership can be defined as ``\,nodes with degree no
smaller than $k$\,'' or ``\,the $r$ best connected nodes\,'', where
rank $r=1...N$ denotes a node's position on the non-increasing degree
list of a graph of size $N$. The rich-club phenomenon is
quantitatively assessed by the metric of rich-club connectivity,
$\varphi$, which is defined as the ratio of the actual number of
links  to the maximum possible number of links among members of a
rich-club. Rich-club connectivity measures how well club members
``\,know'' each other, e.g. $\varphi=1$ means that all the members
have a direct link to any other member, i.e. they form a fully
connected mesh, a clique. The top clique size $n_{clique}$ is defined
as the maximum number of highest rank nodes still forming a clique.

\begin{figure}
\centerline{\psfig{figure=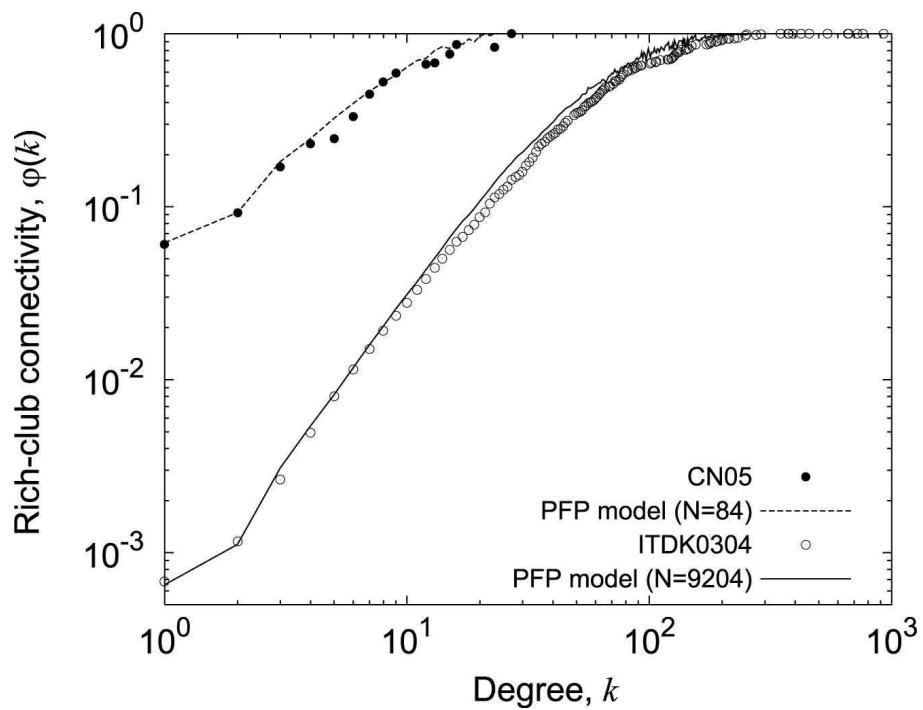,width=12cm}}
\caption{\label{fig:evaluation:rich-degree}Rich-club connectivity vs
degree.}\end{figure}

\begin{figure}[tbh]
\centerline{\psfig{figure=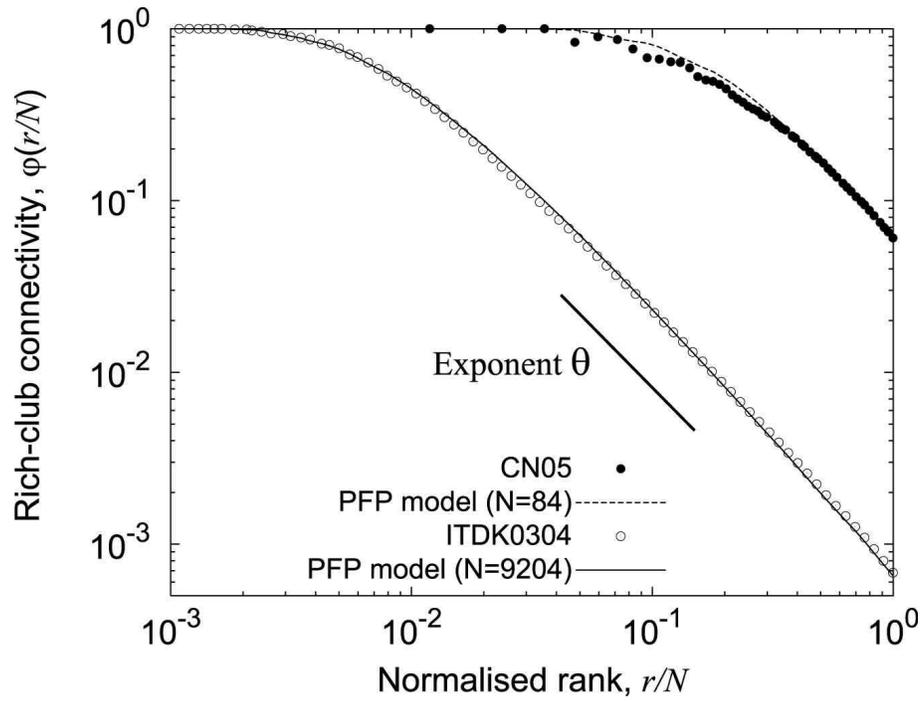,width=12cm}}
\caption{\label{fig:evaluation:rich-rank-normalised}Rich-club
connectivity vs normalized rank.}
\end{figure}

Fig.~\ref{fig:evaluation:rich-degree} illustrates  the rich-club
connectivity as a function of node degree. It shows that a node in
ITDK0304 should have a degree larger than 200 to be a member of
the top clique, whereas a node in CN05 can be a member of the top
clique with a degree of merely 20.
Fig.~\ref{fig:evaluation:rich-rank-normalised} illustrates the
rich-club connectivity as a function of node rank normalized by
the network size $N$. We can see that both CN05 and ITDK0304 obey
a power law of $\varphi(r/N)\sim(r/N)^{\theta}$ with the exponent
$\theta$ of $-1.42$ and $-1.48$ respectively (see
Table~\ref{table:properties}). The PFP networks accurately
resemble all these rich-club properties of the two AS graphs.

\subsubsection{Triangle Coefficient}

Short cycles (e.g.~triangles and quadrangles) encode the redundancy
information in a network structure because the multiplicity of paths
between any two nodes increases with the density of short cycles. The
triangle coefficient, $k_t$, is defined as the number of triangles
that a node shares~\cite{zhou04d}. Different from the widely-studied
clustering coefficient~\cite{watts98}, which can be given as
$c={k_t\over k(k-1)/2}$, the triangle coefficient is able to infer
neighbor clustering information of nodes with different degrees.

\begin{figure}[tbh]
\centerline{\psfig{figure=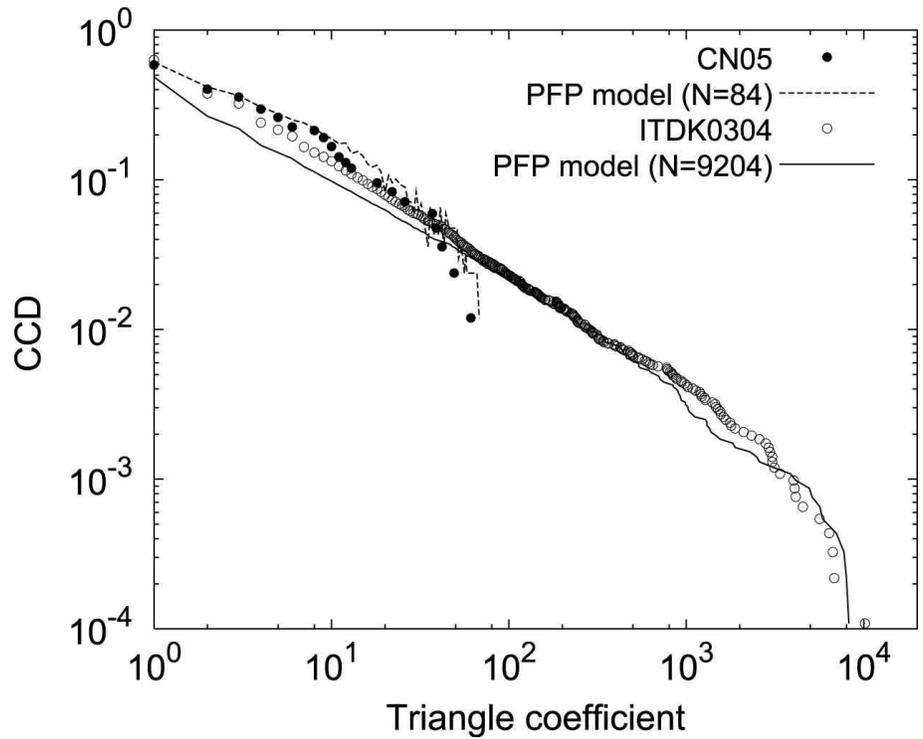,width=12cm}}
\caption{\label{fig:evaluation:PTriangle}Complementary cumulative
distribution (CCD) of triangle coefficient.}\end{figure}

\begin{figure}[tbh]
\centerline{\psfig{figure=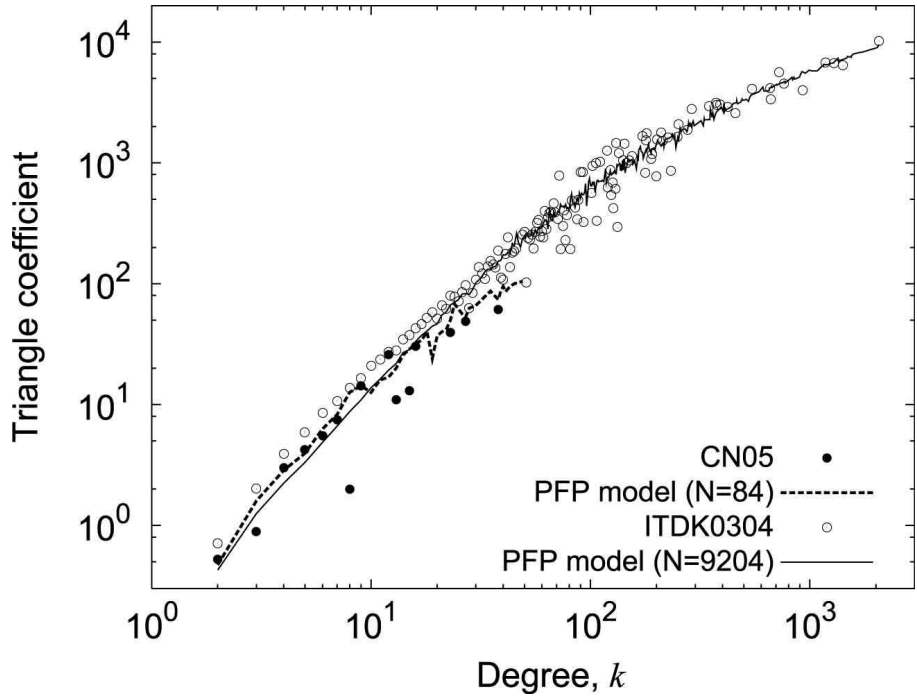,width=12cm}}
\caption{\label{fig:evaluation:Triangle-K} Average triangle
coefficient of $k$-degree nodes.}
\end{figure}

Fig.~\ref{fig:evaluation:PTriangle} shows the complementary
cumulative distribution (CCD) of triangle coefficient. The two AS
graphs exhibit similar power-law distributions.
Fig.~\ref{fig:evaluation:Triangle-K} shows that the correlation
between degree and triangle coefficient of the two AS graphs are
nearly overlapped with each other -- the larger degree a node has,
the more triangles it shares. The PFP networks closely match these
triangular properties of the two AS graphs.

\subsubsection{Shortest Path Length}

The Internet is a small-world network~\cite{watts98} because it is
possible to get to any node via only a few links among adjoining
nodes. The shortest path length $\ell$ is the minimum hop distance
between a pair of nodes. The characteristic path length, $\ell^*$, of
a network is the average of shortest path lengths over all pairs of
nodes. Performance of modern routing algorithms depend strongly on
the distribution of shortest path\,\cite{labovits01}.

\begin{figure}[tbh]
\centerline{\psfig{figure=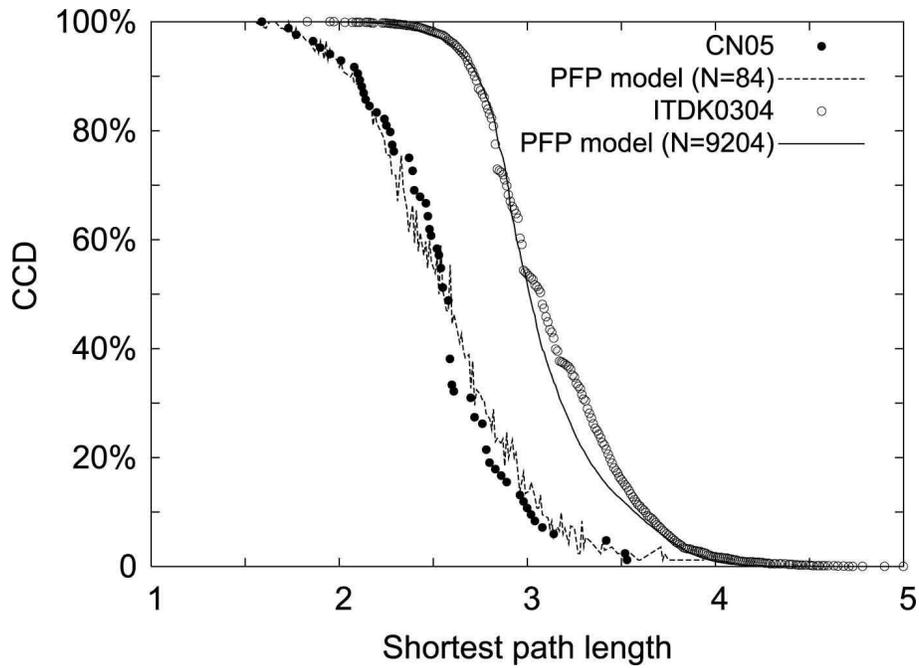,width=12cm}}
\caption{\label{fig:evaluation:PL}Complementary cumulative
distribution (CCD) of shortest path length.}
\end{figure}

\begin{figure}[tbh]
\centerline{\psfig{figure=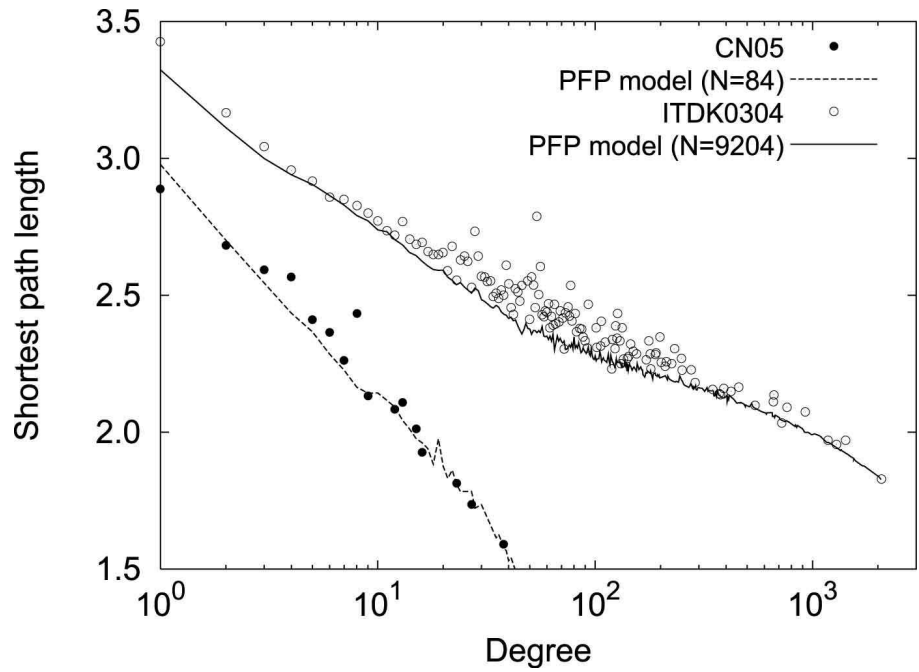,width=12cm}}
\caption{\label{fig:evaluation:L-K} Correlation between shortest path
length and degree.}
\end{figure}

Fig.\,\ref{fig:evaluation:PL} shows the CCD of shortest path length.
It is clear that CN05 is smaller than ITDK0304 (also see $\ell^*$ in
Table~\ref{table:properties}). Fig.\,\ref{fig:evaluation:L-K} shows
that the two AS graphs show a negative correlation between the
shortest path length and node degree, i.e.~in general, the higher
degree a node has, the shorter average distance between the node to
all other nodes. Again the PFP networks precisely resemble the
shortest path properties of the two AS graphs.

\subsubsection{$k$-core Structure}

\begin{figure}[tbh]
\centerline{\psfig{figure=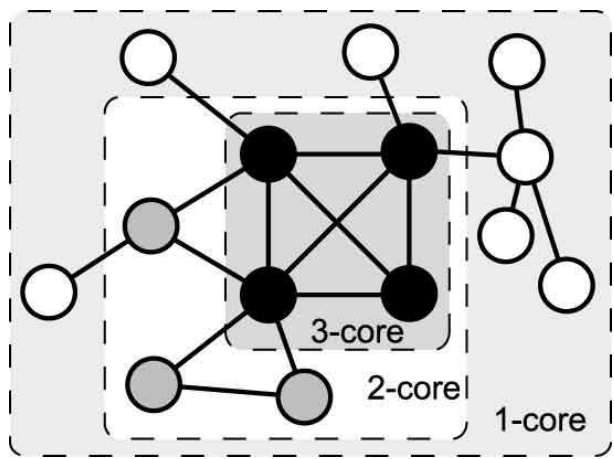,width=8cm}}
\caption{\label{fig:example} Example of $k$-core decomposition.}
\end{figure}

Recently researchers are interested in a topological property called
the $k$-core~\cite{dorogovtsev06}. As shown in
Fig.~\ref{fig:example}, a $k$-core of a network can be obtained by
recursively removing all the nodes of degree less than~$k$, until all
nodes in the remaining graph have at least degree~$k$. A node has the
\emph{coreness} value of $c$, if it belongs to the $c$-core but not
to $(c+1)$-core. Nodes having the same coreness value of $c$ are
called the $c$-shell of a network. The $k$-core decomposition, by
recursively pruning the least connected nodes, disentangles the
hierarchical structure of a network by progressively focusing on its
central cores.

\begin{figure*}[tbh]
\begin{minipage}[t]{8cm}
\centerline{\psfig{figure=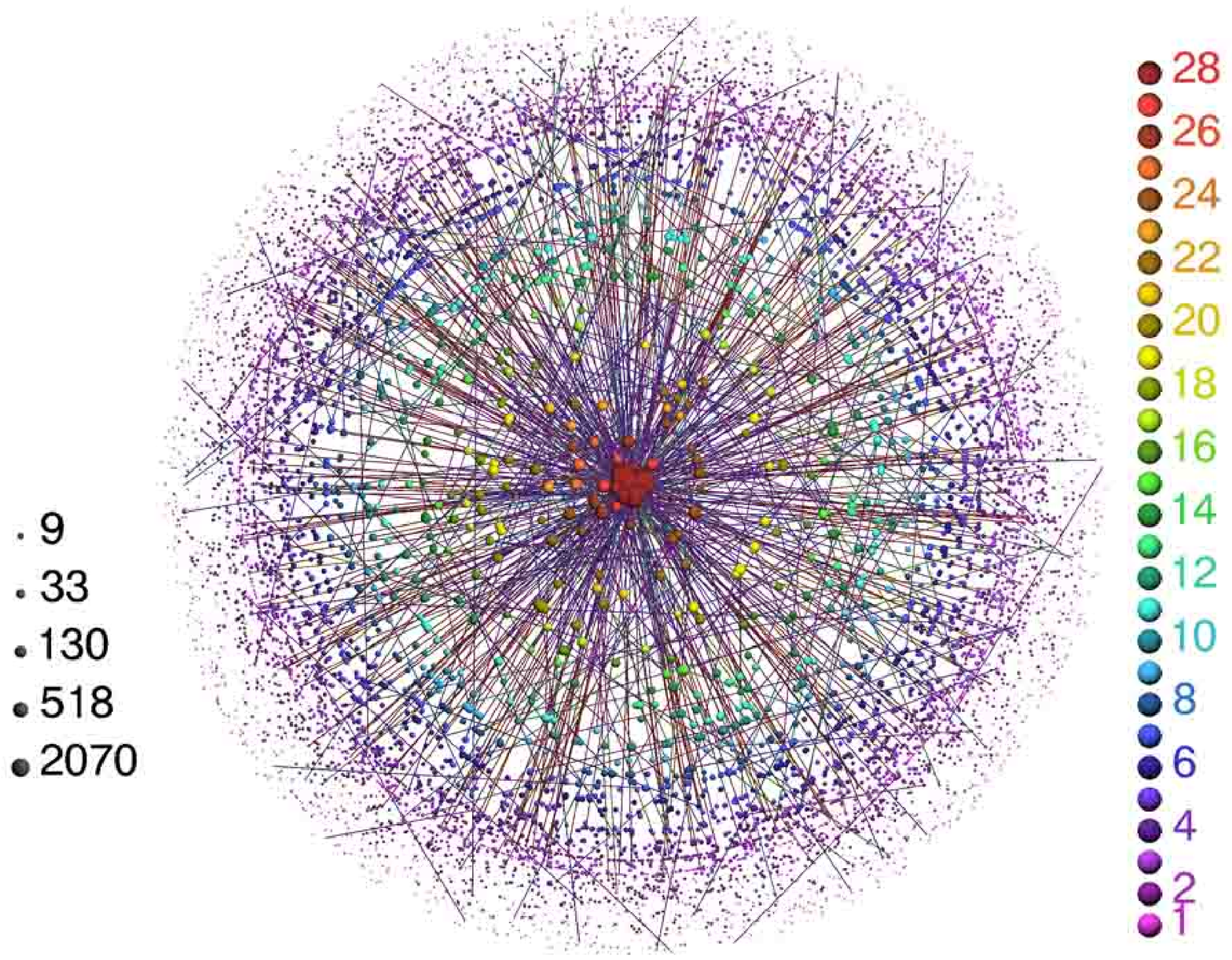,width=8cm}}\centerline{(a).
Internet AS graph (ITDK0304)}
\end{minipage}
\hspace{\fill}
\begin{minipage}[t]{8cm}
\centerline{\psfig{figure=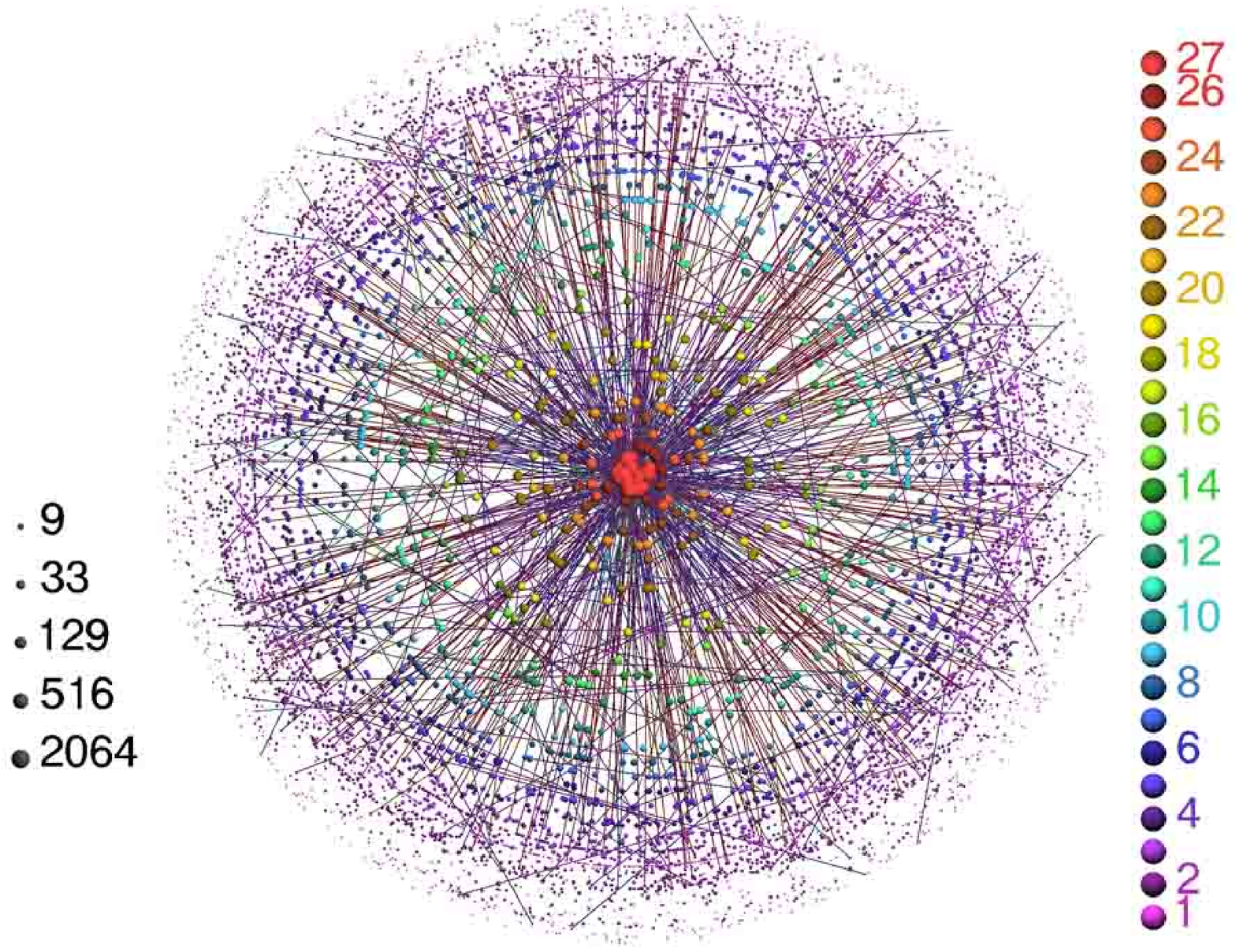,width=8cm}}\centerline{(b).
PFP model (N=9024)}
\end{minipage}
\begin{minipage}[t]{8cm}
\centerline{\psfig{figure=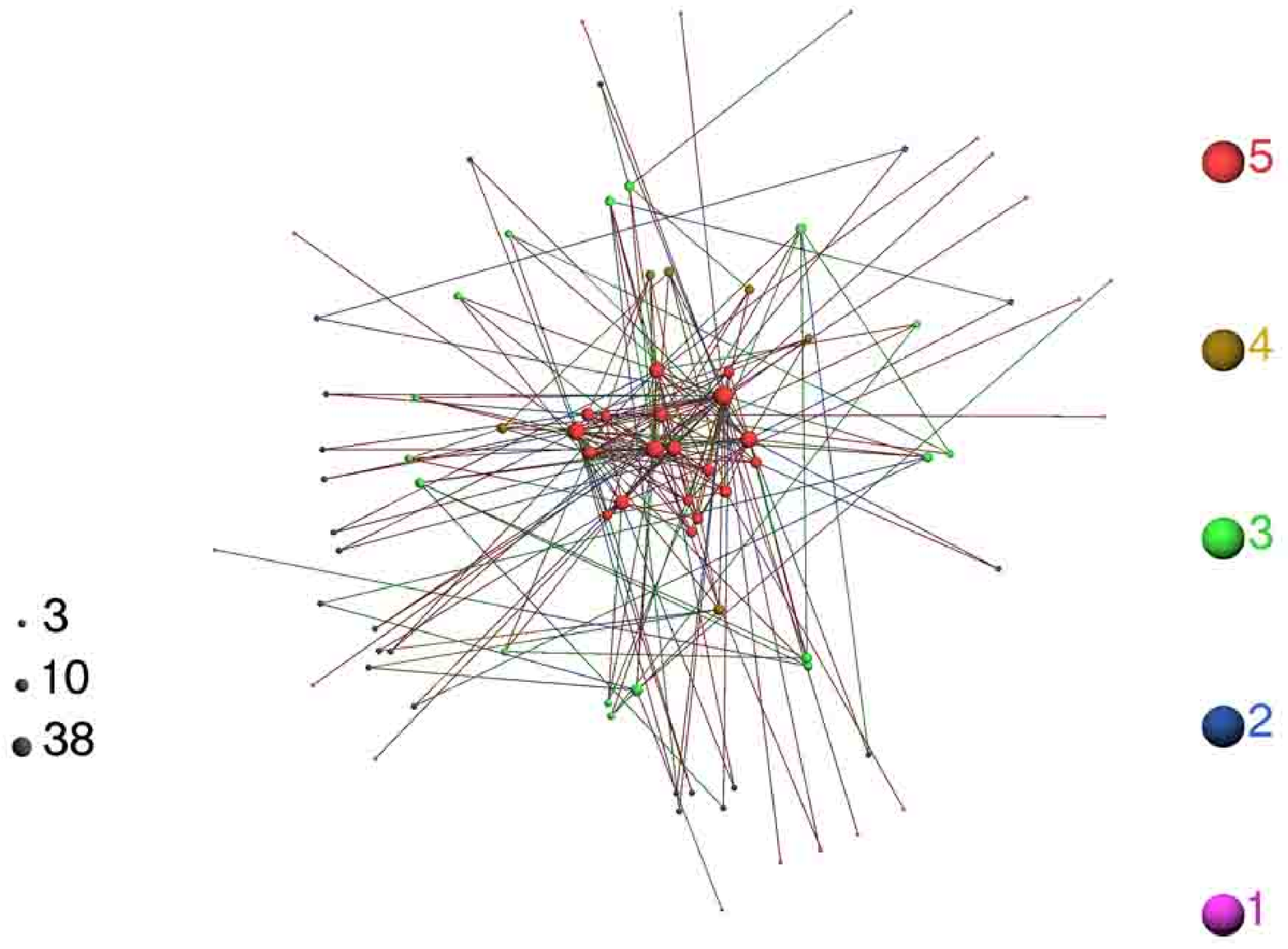,width=8cm}}\centerline{(c).
Chinese Internet AS graph (CN05)}
\end{minipage}
\hspace{\fill}
\begin{minipage}[t]{8cm}
\centerline{\psfig{figure=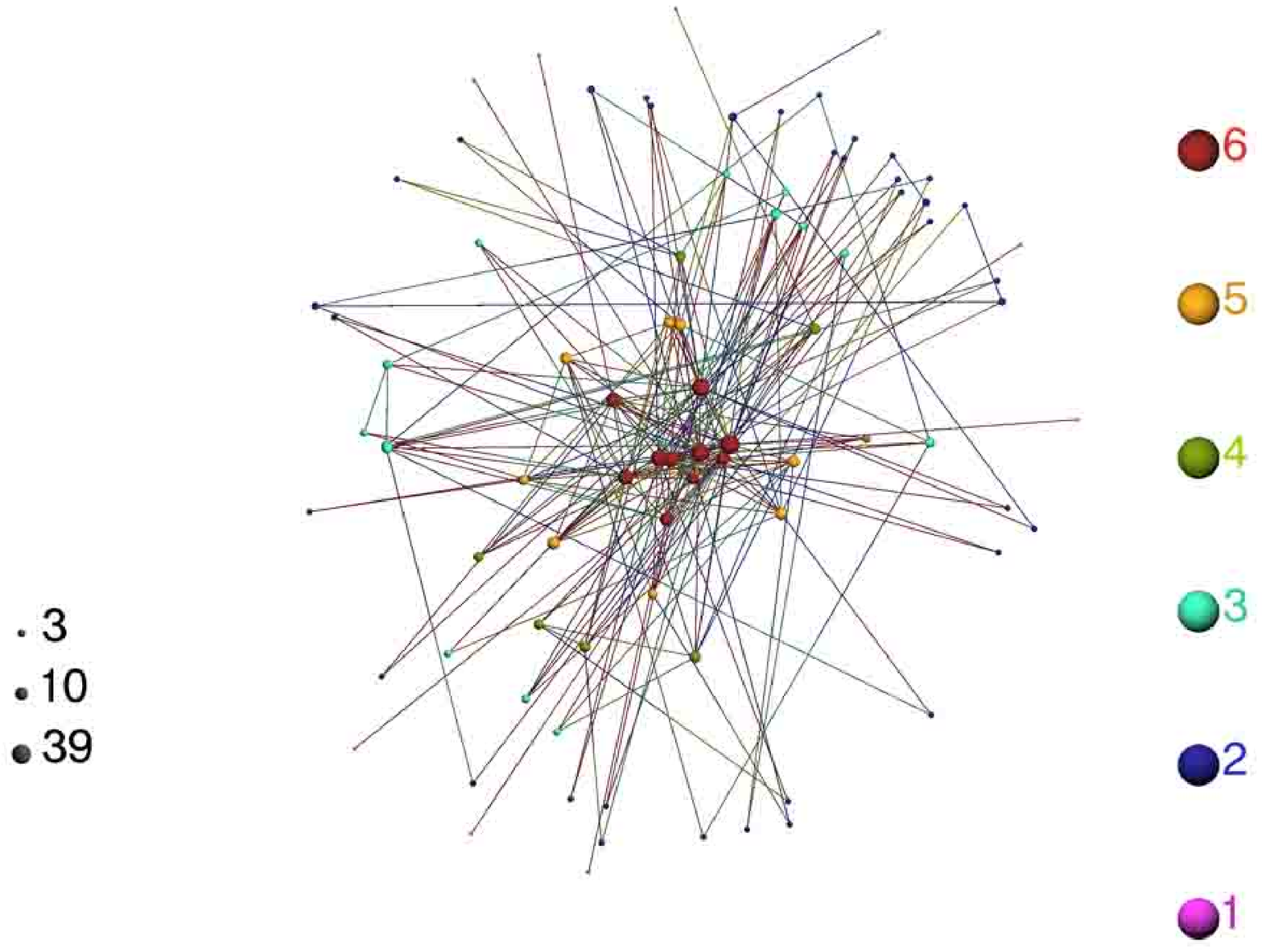,width=8cm}}
\centerline{(d). PFP model (N=84)}
\end{minipage}

\caption{\label{fig:k-core} Visualization of $k$-core structure. }
\end{figure*}

Fig.~\ref{fig:k-core} shows visualizations of the networks $k$-core
structure. Pictures are produced by the tool
LaNet-VI~\cite{alvarez05, lanet-vi}. On each picture a network node
is illustrated as a dot, whose size is exponentially proportional to
the node's degree, and whose color represents the node's coreness
value. Dots are distributed on a series of concentric circular
$c$-shells corresponding to their coreness value~$c$. The diameter of
each shell depends on the shell index $c$, and is proportional to
$c_{max}-c$. The pictures clearly show that the PFP networks closely
resemble the $k$-core structure of the two AS graphs.

\subsection{Discussion}

The above analysis demonstrates that the Chinese Internet AS graph
preserves well the structural characteristics of the global Internet
AS graph. The Internet in China has developed in a social-economic
environment that is characterized by more centralized planning and
less commercial competition than the global Internet dominated by the
West. However this does not make the evolution of the Internet
AS-level topology in China different from the West, because the local
social-economic factors mainly affect the development of the
last-mile access networks, e.g.~China has a small number of IPSs, but
all ISPs are very large. Whereas the evolution of the Internet
AS-level topology is primarily influenced by technological factors,
which are fairly universal. That is to say, when making a decision on
a BGP peering relationship~\cite{chang06}, all ISPs around the world
tend to consider the same technical issues, such as performance
objectives and technical constrains. In our future work, we will
investigate the Internet AS-level subgraphs in other countries and
regions.

\section{Conclusion}

In this paper we present the first traceroute measurement of the
Chinese Internet AS-level topology collected from probing servers
inside the country. We compare the global Internet AS graph with the
Chinese Internet AS graph using the PFP model as a yardstick. We
examine a diversified set of topological characteristics of the two
AS graphs including degree distribution, degree correlation,
rich-club connectivity, triangle coefficient, shortest path length
and $k$-core structure. Remarkably the same parameters in the PFP
model to describe the global Internet AS graph also describe the
Chinese Internet AS graph. This is the first demonstration of the
Internet's topological fractality, or self-similarity, performed at
the level of topology evolution modeling.

\section{Acknowledgments}

This work is partly supported by the UK Nuffield Foundation under
Grant No.~NAL/01125/G and the National Natural Science Foundation of
China under Grant No.~60673168. The Internet data kit ITDK0304 is
provided by CAIDA.


\newpage

\section*{Figure captions}

\begin{enumerate}

\item Interactive growth of the PFP model.

\item Degree distribution.

\item Nearest-neighbors average degree of $k$-degree nodes.

\item Rich-club connectivity vs degree.

\item Rich-club connectivity vs normalized rank.

\item Complementary cumulative distribution (CCD) of triangle
coefficient.

\item Average triangle coefficient of $k$-degree nodes.

\item Complementary cumulative distribution (CCD) of shortest path
length.

\item Correlation between shortest path length and degree.

\item Example of $k$-core decomposition.

\item Visualization of $k$-core structure.

\end{enumerate}

\newpage

\end{document}